\documentclass[a4paper,11pt]{article}
\usepackage{pos}

\newcommand{\lb}{\left(}
\newcommand{\rb}{\right)}
\newcommand{\GeV}{{\ensuremath\rm GeV}}
\newcommand{\TeV}{{\ensuremath\rm TeV}}
\newcommand{\eqn}{equation}
\newcommand{\lam}{\lambda}

\title{Extended scalar sectors from all angles
- Mostly at lepton colliders -}

\author*[a,b]{Tania Robens}

\affiliation[a]{Institute Rudjer Boskovic,\\
  Bijenicka cesta 54, 10000 Zagreb, Croatia}
\affiliation[b]{Theoretical Physics Department, CERN,\\
 1211 Geneva 23, Switzerland}

\emailAdd{trobens@irb.hr}

\abstract{I briefly discuss the current state of the art for models with extended scalar sectors and give examples for corresponding investigations, with a focus on processes at lepton colliders.\\
RBI-ThPhys-2025-20, CERN-TH-2025-082}

\FullConference{Proceedings of the Corfu Summer Institute 2024 "School and Workshops on Elementary Particle Physics and Gravity" (CORFU2024)\
12 - 26 May, and 25 August - 27 September, 2024\\
Corfu, Greece\\}

\begin{document}
\maketitle

\section{Introduction}

After the discovery of the Higgs boson by the LHC experiments \cite{ATLAS:2012yve,CMS:2012qbp}, particle physics has entered in exciting era. A crucial question is whether the boson discovered by the LHC experiments is the only scalar boson realized in nature, or whether it is part of a more extended scalar sector that could in principle explain additional current puzzles of nature as dark matter or the flavour structure.
Although the Standard Model (SM) parameters by themselves are all determined, not all couplings have been measured yet, as e.g. triple and quartic Higgs couplings. This as well as theoretical and experimental uncertainties still leaves room for a large variety of new physics scenarios. The latter could manifest themselves via either direct deviations in e.g. invariant mass distributions, signatures that are not present in the SM alone (as e.g. events with large missing transverse energy), or deviations in SM-like quantities that are small and therefore can be a result of loop induced new physics effects.

In this work, I briefly discuss a short overview on various model with extended scalar sectors, with a focus on models that can be tested at future lepton colliders. I also briefly mention the recasting of various searches for new physics scenario within the Inert Doublet Model.
\section{Inert Doublet Model: Recast}
The Inert Doublet Model (IDM) \cite{Deshpande:1977rw,Barbieri:2006dq,Cao:2007rm} is a two Higgs doublet model with an exact $\mathbb{Z}_2$ symmetry that renders the lightest scalar of the second doublet stable and therefore provides a dark matter candidate. Furthermore, the symmetry prevents the coupling to fermions of that doublet, such that the additional bosons decay via electroweak gauge bosons. The potential is given by

The scalar potential of IDM is given by 
\begin{align}
    V_\text{IDM} =&\ \mu_1^2 |\Phi_1|^2 + \mu_2^2 |\Phi_2|^2 + \frac{1}{2} \lambda_1 |\Phi_1|^4 + \frac{1}{2} \lambda_2 |\Phi_2|^4 + \lambda_3 |\Phi_1|^2 |\Phi_2|^2 + \lambda_4 |\Phi_1^\dagger \Phi_2|^2 \nonumber\\
    &+ \frac{1}{2} \lambda_5 \left[(\Phi_1^\dagger\Phi_2)^2+\text{h.c.}\right], 
\end{align}
with
\begin{align}
    \Phi_1=\begin{pmatrix}
      G^+\\
      \frac{1}{\sqrt{2}}(v+h+iG^0)
    \end{pmatrix}\,,
    \quad \text{and} \quad
    \Phi_2=\begin{pmatrix}
      H^+\\
      \frac{1}{\sqrt{2}}(H+iA)
    \end{pmatrix}\,,
\end{align}
and where $h\,H,\,A,\,H^+$ denote the physical mass eigenstates.

In this work we follow the scan presented in \cite{Ilnicka:2015jba}  and subsequently updated in \cite{Kalinowski:2018ylg,Dercks:2018wch,Kalinowski:2020rmb,Braathen:2024lyl}.

At the LHC, a prime signature of the model are processes with electroweak gauge bosons and missing transverse energy, leading typically to multi-lepton signatures. An overview on possible cross sections for such processes at various collider stages has been provided in \cite{Kalinowski:2020rmb} . Similar final states are also accessible in other new physics scenarios, such as supersymmetric setups or other scalar extensions with dark matter candidates. Therefore, a natural question is how sensitive searches that were designed for other models are for the IDM, or whether it is necessary to design search strategies targeted at this model specifically.

Recasts for the IDM for such signatures have already been presented in the literature, see e.g. \cite{Belanger:2015kga} or \cite{Belyaev:2022wrn} for recent work. However, a large part of the parameter space the searches might be sensitive to is already constrained by e.g. results from dark matter as e.g. relic density or direct detection constraints.

\begin{center}
\begin{figure}
\begin{center}
  \begin{minipage}{0.35\textwidth}
    \includegraphics[width=\textwidth]{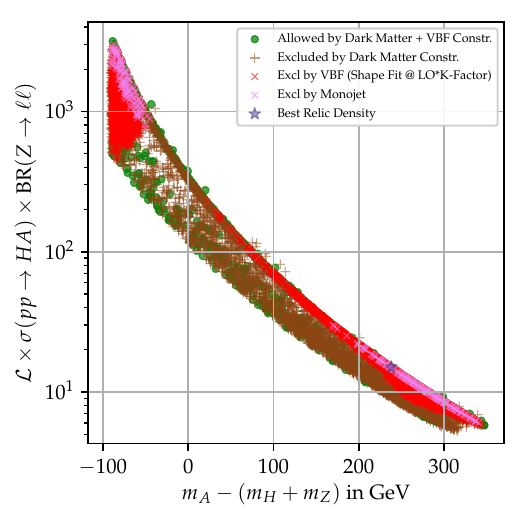}
  \end{minipage}
    \begin{minipage}{0.35\textwidth}
      \includegraphics[width=\textwidth]{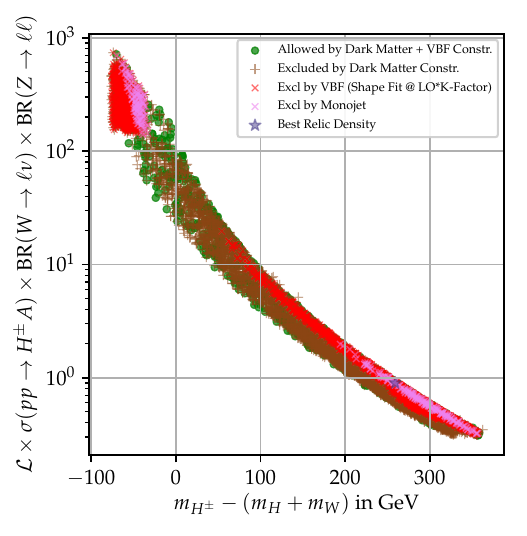}
    \end{minipage}
\caption{\label{fig:danres} Total rates as a function of the available phase space using a mass estimation for $2\,\ell$ and $3\,\ell$ lepton searches for the IDM. High rates are correlated with small available phase space, in turn leading to suppression for high missing transverse energy cuts. Taken from \cite{Dercks:2018wch}.}
\end{center}
\end{figure}
\end{center}

In \cite{Dercks:2018wch}, we investigated constraints on the IDM stemming from recasting the searches for $h\,\rightarrow\,\text{invisible}$ from early run 2 LHC data. As a side product, we additionally checked whether any of the points that we investigated might be excluded by multi-lepton searches with missing transverse energy that were designed for supersymmetric scenarios \cite{ATLAS:2017nyv,ATLAS:2018ojr}. We found that the relatively high cuts on missing transverse energy of around 100 \GeV~ basically cut out all regions of parameter space with higher cross sections. A rough estimate of this dependence is shown in figure \ref{fig:danres}, where we estimate the available phase space based on kinematic properties of the parameter points. We observe there is a direct correlation between high cross sections and small expected missing transverse energy\footnote{In principle more detailed studies would be needed here. The argument leading to the above plot can only be seen as a first estimate.}.

As a follow up, we then investigated the constraints that come from a more recent search for dilepton and missing energy, primarily designed for the THDMa, a two Higgs doublet model with an additional pseudoscalar field that serves as a portal to the dark sector \cite{Ipek:2014gua,No:2015xqa,Goncalves:2016iyg,Bauer:2017ota,Tunney:2017yfp,Pani:2017qyd,LHCDarkMatterWorkingGroup:2018ufk,Haisch:2018kqx,Abe:2018emu,Haisch:2018hbm,Haisch:2018bby,Abe:2019wjw,Butterworth:2020vnb,Arcadi:2020gge,Argyropoulos:2021sav,Robens:2021lov,Arcadi:2022dmt,Arcadi:2022lpp,Haisch:2023rqs,Argyropoulos:2024yxo}. This search has been shown to put relatively strong constraints in that model (see e.g. \cite{Robens:2021lov}), and it is therefore natural to investigate whether this can pose constraints on the IDM.

\begin{center}
\begin{figure}
\begin{center}
\begin{minipage}{0.49\textwidth}
\begin{center}
\includegraphics[width=\textwidth]{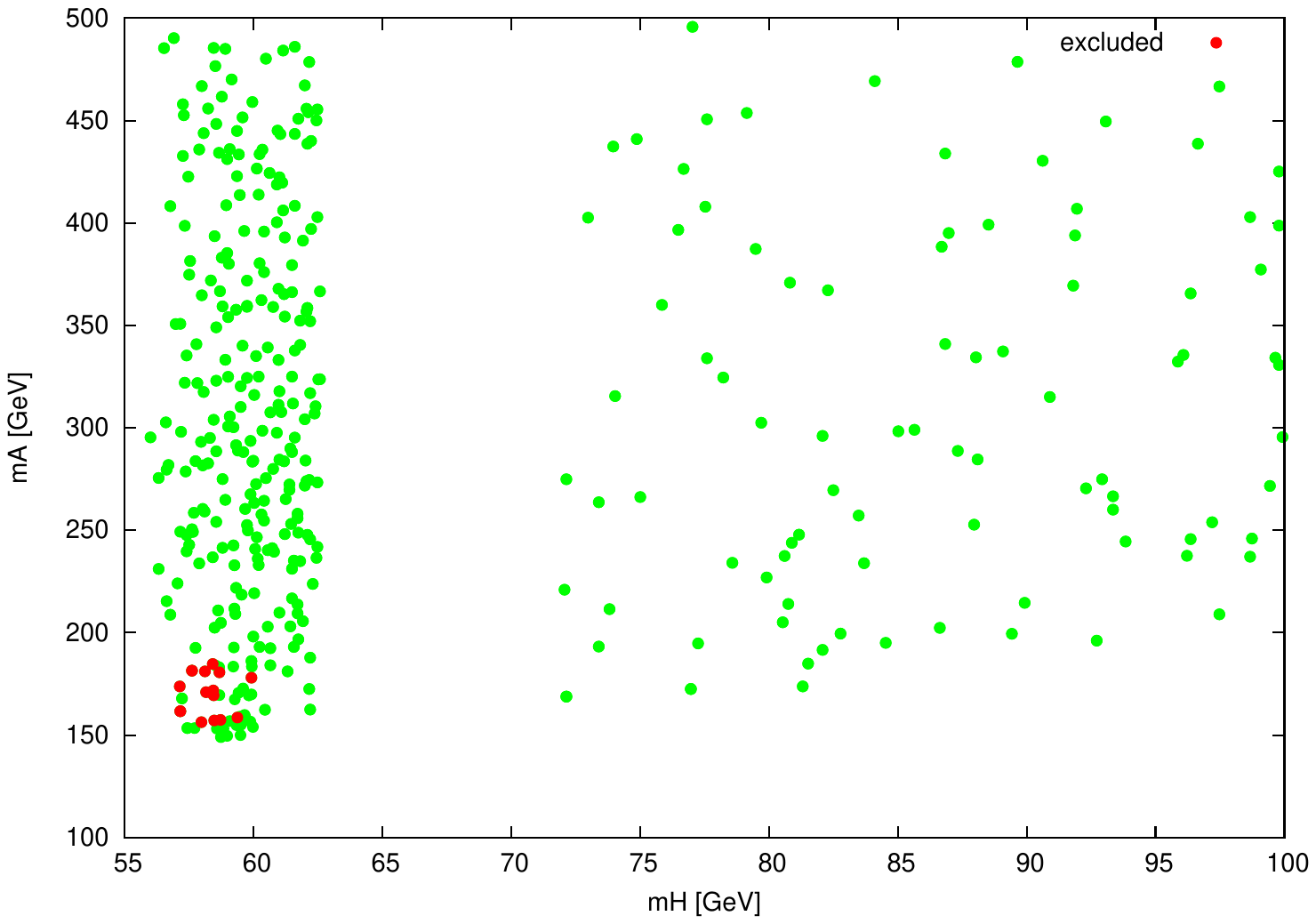}\\
\end{center}
\end{minipage}
\begin{minipage}{0.49\textwidth}
\begin{center}
\includegraphics[width=\textwidth]{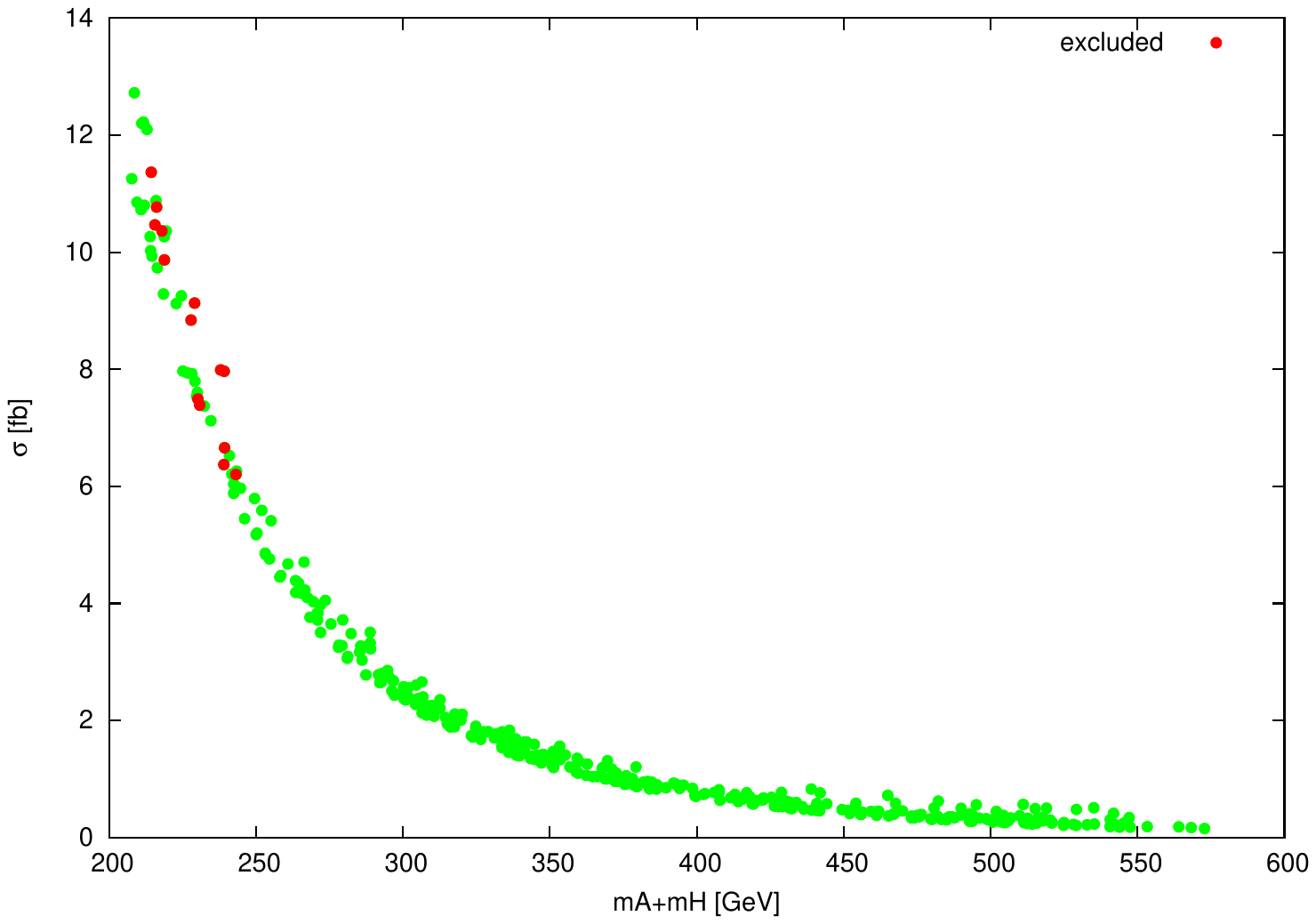}\\
\end{center}
\end{minipage}
\end{center}
\caption{\label{fig:idmrecast} Allowed and excluded points from a scan presented in \cite{chrisjayitame}, using the recasted THDMa search as implemented in CheckMate. {\sl Left:}  $\lb m_H,\,m_A\rb$ plane. {\sl Right:} $\lb m_H+m_A,\sigma_{HA}\rb$ plane. Very few points are sensitive.}
\end{figure}
\end{center}

In ongoing work \cite{chrisjayitame}, we therefore have used the publicly available code CheckMATE \cite{Drees:2013wra,Dercks:2016npn} to recast the corresponding search \cite{ATLAS:2021gcn} in the IDM. We show preliminary results of this study in figure \ref{fig:idmrecast}, with allowed and excluded points both in the $\lb m_H,\,m_A\rb$ and $\lb m_H+m_A,\sigma \rb$ plane.
We see that in principle only a small region of the parameter space is sensitive to this search. In addition, as various channels are contributing to the corresponding final state, one cannot draw clear exclusion contours in the di-mass plane as naively expected.
 
\begin{center}
\begin{figure}
\begin{center}
\includegraphics[width=0.7\textwidth]{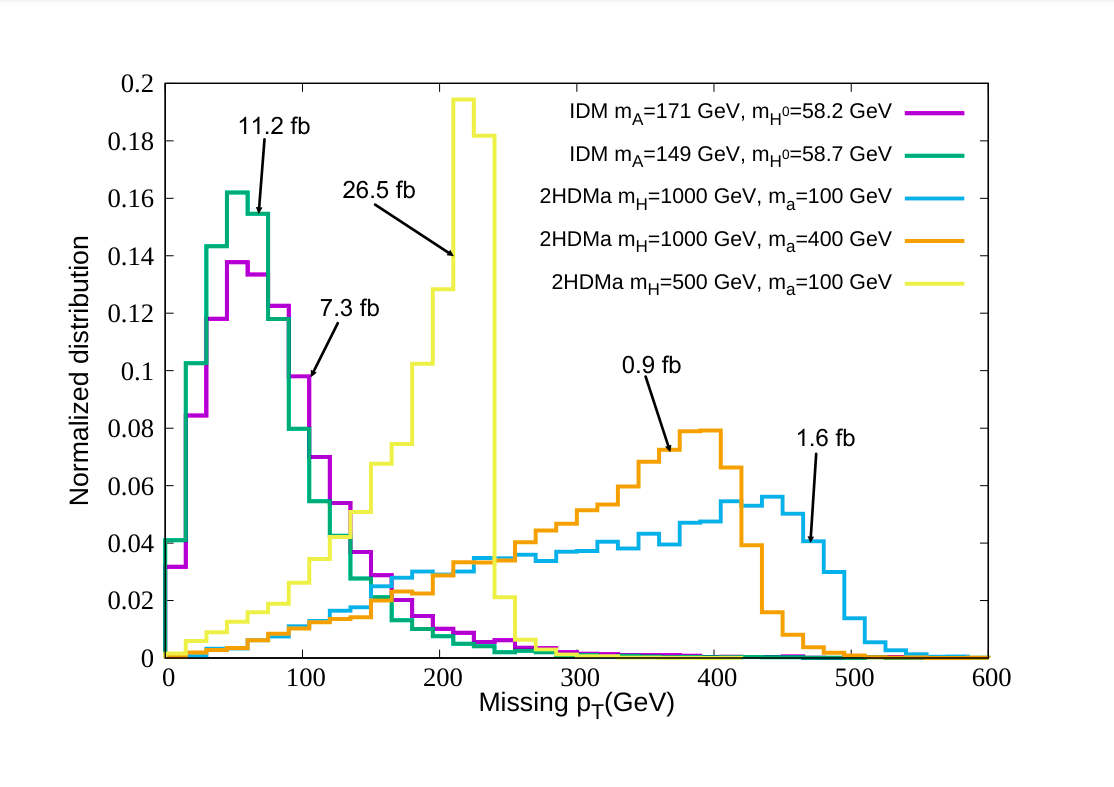}
\end{center}
\caption{\label{fig:kine} Transverse missing momentum distributions for various parameter points in the IDM and THDMa, including cross section predictions. Although rates for the THDMa points are significantly lower, all of these are forbidden by the search, while the IDM points are still allowed. This is mainly due to the applied cut in missing transverse momentum.}
\end{figure}
\end{center}

It is interesting to see that for the THDMa, parameter points with much smaller cross sections are already excluded. We investigate the reason for this behaviour in figure \ref{fig:kine}, which shows the missing energy distribution for various new physics points in the IDM and THDMa, respectively, in addition to the corresponding production cross sections. We see that the IDM points have a peak toward lower missing transverse momentum, while the THDMa points peak at higher values. As before, large constraints on missing transverse momentum cut out the interesting regions in parameter space for the IDM. 

\section{New physics at Higgs factories}
In both the last European strategy (see e.g. \cite{EuropeanStrategyforParticlePhysicsPreparatoryGroup:2019qin} for the Physics Briefing Book) as well as the Snowmass/ P5 process (see e.g. \cite{p5rep}), a Higgs factory was identified as one of the next important world wide collider projects. It is therefore interesting to investigate in which way novel scalars could be produced and discovered at such a machine.

For Higgs-like particles, one of the main production modes is the so-called scalar strahlung, leading to

\begin{\eqn}\label{eqn:zstrahl}
e^+\,e^-\,\rightarrow\,Z\,h_i
\end{\eqn}
via offshell Zs in the s-channel, and where $h_i$ denotes a CP even neutral scalar. Another production mode would be vector boson fusion (VBF) leading to the final state
\begin{\eqn*}
e^+\,e^-\,\rightarrow\,\nu\,\bar{\nu}\,h_i.
\end{\eqn*}

\begin{center}
\begin{figure}
\begin{center}
\includegraphics[width=0.5\textwidth]{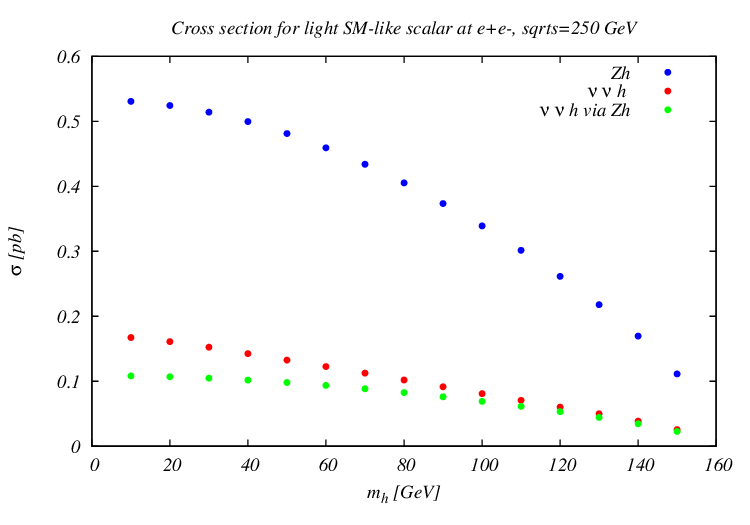}
\end{center}
\caption{\label{fig:xsecs_250} Production cross sections of an SM like scalar in various production processes, at a Higgs factory with a center-of-mass energy of 250 \GeV.
Taken from \cite{Robens:2024wbw}.}
\end{figure}
\end{center}

Note that the process specified in eqn (\ref{eqn:zstrahl}) also contains the latter process. We therefore plot the production cross section for a SM-like scalar in figure \ref{fig:xsecs_250} for a center-of-mass energy of 250 \GeV. The production cross sections were calculated using Madgraph \cite{Alwall:2011uj} at leading order.
We see that, in particular for masses $\geq\,80\,\GeV$, the $Z\,h$ production is dominant, and we therefore concentrate on this production channel for the rest of this section. 

\begin{center}
\begin{figure}
\begin{center}
\includegraphics[width=0.49\textwidth]{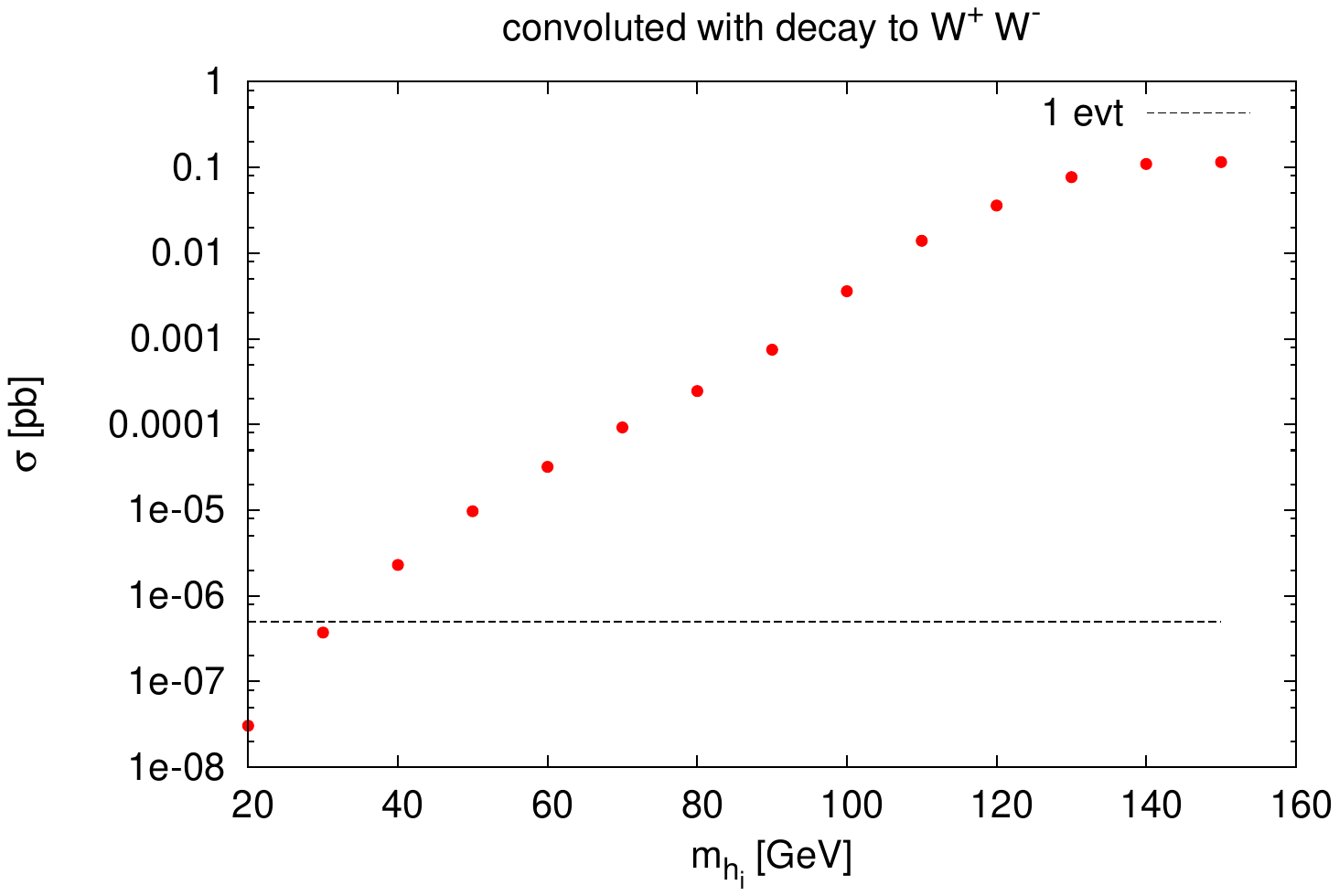}
\includegraphics[width=0.49\textwidth]{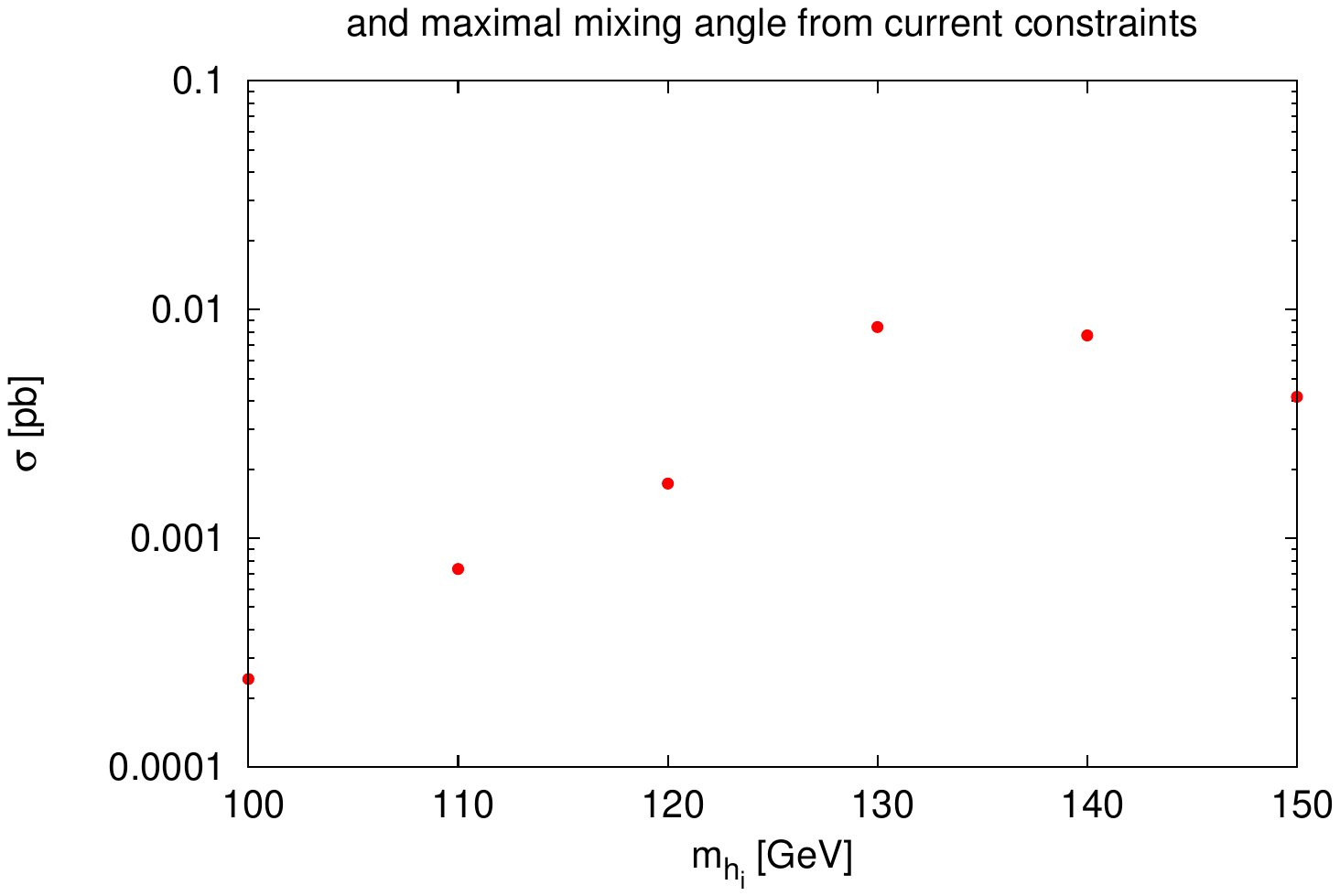}
\caption{\label{fig:convu} Production rates for a Higgs-like scalar at a 250 \GeV~ Higgs factory: {\sl (left)} convoluted with branching ratios to $W^+\,W^-$ final states and {\sl (right)} with realistic production rescaling as e.g. discussed in \cite{Feuerstake:2024uxs}.}
\end{center}
\end{figure}
\end{center}

In \cite{deBlas:2024bmz}, it was pointed out that one channel that has not been investigated yet in phenomenological studies was
\begin{\eqn*}
e^+\,e^-\,\rightarrow\,W^+\,W^-\,Z,
\end{\eqn*}
that can partially be mediated by an additional scalar from eqn (\ref{eqn:zstrahl}) decaying into a $W^+\,W^-$ pair. In order to determine the best parameter point, the production cross sections in figure \ref{fig:xsecs_250} need to be convoluted with the respective branching ratios for that final state as well as realistic production rates. We here refer to a scenario where the extra scalar can e.g. stem from a singlet extension, this way obeying the sum rule \cite{Gunion:1990kf}
\begin{\eqn*}
\sum_i\,g^2_i\,(h_i)\,=\,g^2_{SM}
\end{\eqn*}
where the $g_i$ denote the couplings to vector bosons for additional new scalars in the model and $g$ is the respective coupling in the SM. The current measurements of the Higgs signal strength this way already put strong constraints on the allowed parameter space (see e.g. \cite{Feuerstake:2024uxs} for a recent update). In figure \ref{fig:convu}, the left hand side shows the rate convoluted with the branching ratio to $W^+\,W^-$, while the right hand side additionally takes into account realistic current constraints on the production rescaling derived using HiggsTools \cite{Bahl:2022igd} based on HiggsBounds \cite{Bechtle:2008jh,Bechtle:2011sb,Bechtle:2013wla,Bechtle:2020pkv} and HiggsSignals \cite{Bechtle:2013xfa,Bechtle:2014ewa,Bechtle:2020uwn}. We see that, after taking these additional factors into account, the largest rates stem from new scalar masses around $130-140\,\GeV$. In \cite{filipcorfu}, some preliminary results on studies for these parameter points are presented.

\section{New physics at Muon Colliders}
Another interesting point is the investigation of the potential of high energy lepton colliders, like e.g. colliders with muonic initial states, that have recently attracted large interest (see e.g. \cite{Black:2022cth,Accettura:2023ked,InternationalMuonCollider:2024jyv} for recent community reports). One advantage of such high-energy machines is that they experience an enhancement of vector boson fusion processes as the masses of the emitted electroweak gauge bosons become small with respect to the available collider energy, therefore leading to a log-type enhancement of the respective cross sections. In particular, this means that processes that are dominated by vector boson fusion channels are largely enhanced, see e.g. \cite{Dawson:1984gx,Kane:1984bb,Chanowitz:1984ne,Gunion:1986gm,Chen:2016wkt,Kunszt:1987tk,Costantini:2020stv,Han:2020uid,Ruiz:2021tdt,Garosi:2023bvq,Denner:2024yut}.

\begin{figure}[t!]
\begin{center}
    \includegraphics[width=0.7\textwidth]{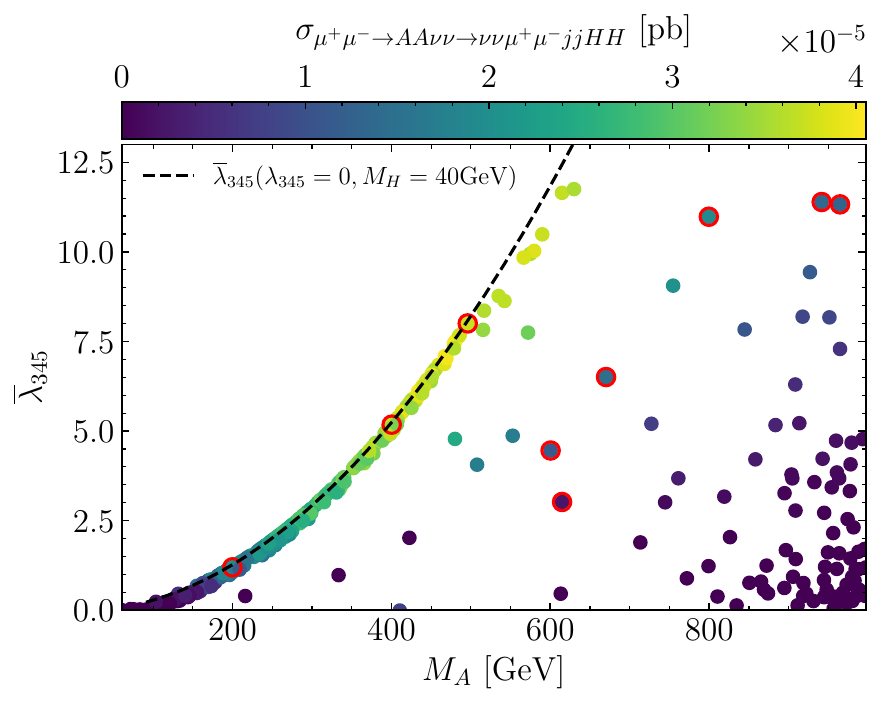}
    \caption{
    Total cross-section, with pre-cuts, for the channel given in eqn (\ref{eq:channel}) at $10$~TeV in the plane of $M_A$ and $\bar{\lambda}_{345}$. The dashed line corresponds to the maximal possible value of $\bar{\lambda}_{345}$ as a function of $M_A$, for $\lambda_{345}=0$ and $M_H=40\ \GeV$. Points lying on or near the dashed line come from the parameter scan with maximal $\bar{\lam}_{345}$ coupling. Red circled points mark specific benchmark points. Taken from \cite{Braathen:2024lyl}.
    \label{fig:xsecs_mumu}}
    \end{center}
\end{figure}

In \cite{Braathen:2024lyl}, we have investigated the pair-production of $A\,A$ in the IDM at a muon collider in the vector boson fusion topology, focusing on a machine with a center-of-mass energy of 10 \TeV. This paper equally contains the newest updates to the parameter space of the IDM.  A crucial quantity in this channel is given by
\begin{\eqn*}
\bar{\lam}_{345}\,=\,\lam_{345}\,+\,\frac{2\,\lb M_A^2-M_H^2\rb}{v^2},
\end{\eqn*}
where $\lam_{345}\,\equiv\,\lam_3+\lam_4+\lam_5$, $M_{A(H)}$ are the masses of the $A\,(H)$ and $v$ is the vacuum expectation value of the SM-like doublet. This coupling determines the rates of the s-channel contributions to $A\,A$ production in VBF. In figure \ref{fig:xsecs_mumu},  we show the production cross sections at a 10 \TeV~ muon collider for the process

\begin{equation}
    \label{eq:channel}
    \mu^+ \mu^- \rightarrow \nu_\mu \bar{\nu}_\mu A A \rightarrow \nu_\mu \bar{\nu}_\mu j j \ell^+ \ell^- H H,
\end{equation}

in the $\lb M_A,\,\bar{\lam}_{345} \rb$ plane, where we equally display the upper limit on $\bar{\lam}_{345}$ for $\lam_{345}\,=\,0,\,M_H\,=\,40\,\GeV$, where the latter values are motivated by dark matter constraints. We see that largest production rates are obtained for maximal $\bar{\lam}_{345}$ values, together with a relatively light mass for $A$. 

\begin{figure}[tb!]
    \centering
    \includegraphics[width=.49\textwidth]{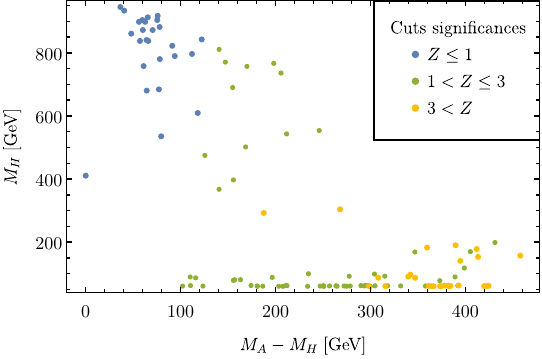}
    \includegraphics[width=.49\textwidth]{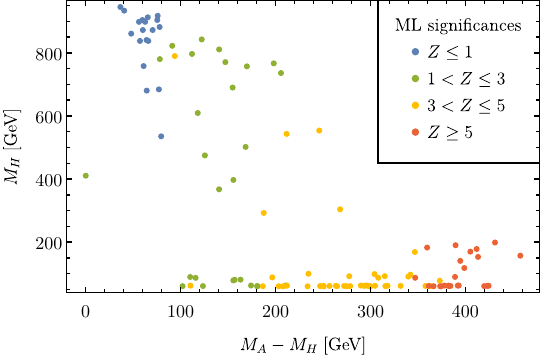}
    \caption{Subset of parameter points that were used for the cut-and-count analysis (left) and the ML analysis (right) shown on the $M_A - M_H$ vs.~$M_H$ plane. Figure taken from \cite{Braathen:2024lyl}.}
    \label{fig:full_scan_signif}
\end{figure}

A dedicated search strategy including background simulation and cut as well as machine learning based setups has been discussed in \cite{Braathen:2024lyl} and will not be repeated here. We show the results for both cut and machine learning based significances in figure \ref{fig:full_scan_signif}, where we define the significance according to \cite{Cowan:2010js}

\begin{\eqn*}
Z = \sqrt{2 \left[ (S + B) \ln\left( 1 + \frac{S}{B} \right) - S \right]}
\end{\eqn*}

In general, we find that machine based techniques supersede the cut based searches as presented in the above work, with significances larger than 5 being possible for relatively light dark matter masses $\sim\,200\,\GeV$ and large mass differences allowing for large $\bar{\lam}_{345}$.

\section{Conclusion}
In this work, I gave a brief overview on work that discussed new physics scenarios with extended scalar sectors at current and future colliders. I focussed on the importance of designing dedicated searches for new physics scenarios with varying topologies, by means of recasting the dilepton search for the THDMa for the Inert Doublet Model. Furthermore, I gave a brief overview on some aspects of loght scalar searches at Higgs factories as well as a discussion of a search for the Inert Doublet Model at future muon colliders. In general, these three topics can only give a glance on all possible scenarios that relate to extended scalar sectors, with a wide variety of additional topics that cannot be covered in this short proceeding contribution.
\section*{Acknowledgements}
This work was supported by the National Science Centre, Poland, under the OPUS research project no. 2021/43/B/ST2/01778. TR is furthermore supported by the Croatian Science Foundation (HRZZ) under project IP-2022-10-2520.

\bibliography{lit}

\begin{thebibliography}{10}

\bibitem{ATLAS:2012yve}
ATLAS, G.~Aad {\em et~al.},
\newblock Phys. Lett. B {\bf 716}, 1 (2012), 1207.7214.

\bibitem{CMS:2012qbp}
CMS, S.~Chatrchyan {\em et~al.},
\newblock Phys. Lett. B {\bf 716}, 30 (2012), 1207.7235.

\bibitem{Deshpande:1977rw}
N.~G. Deshpande and E.~Ma,
\newblock Phys. Rev. D {\bf 18}, 2574 (1978).

\bibitem{Barbieri:2006dq}
R.~Barbieri, L.~J. Hall, and V.~S. Rychkov,
\newblock Phys. Rev. D {\bf 74}, 015007 (2006), hep-ph/0603188.

\bibitem{Cao:2007rm}
Q.-H. Cao, E.~Ma, and G.~Rajasekaran,
\newblock Phys. Rev. D {\bf 76}, 095011 (2007), 0708.2939.

\bibitem{Ilnicka:2015jba}
A.~Ilnicka, M.~Krawczyk, and T.~Robens,
\newblock Phys. Rev. D {\bf 93}, 055026 (2016), 1508.01671.

\bibitem{Kalinowski:2018ylg}
J.~Kalinowski, W.~Kotlarski, T.~Robens, D.~Sokolowska, and A.~F. Zarnecki,
\newblock JHEP {\bf 12}, 081 (2018), 1809.07712.

\bibitem{Dercks:2018wch}
D.~Dercks and T.~Robens,
\newblock Eur. Phys. J. C {\bf 79}, 924 (2019), 1812.07913.

\bibitem{Kalinowski:2020rmb}
J.~Kalinowski, T.~Robens, D.~Sokolowska, and A.~F. Zarnecki,
\newblock Symmetry {\bf 13}, 991 (2021), 2012.14818.

\bibitem{Braathen:2024lyl}
J.~Braathen, M.~Gabelmann, T.~Robens, and P.~Stylianou,
\newblock (2024), 2411.13729.

\bibitem{Belanger:2015kga}
G.~Belanger {\em et~al.},
\newblock Phys. Rev. D {\bf 91}, 115011 (2015), 1503.07367.

\bibitem{Belyaev:2022wrn}
A.~Belyaev, U.~Blumenschein, A.~Freegard, S.~Moretti, and D.~Sengupta,
\newblock JHEP {\bf 09}, 173 (2022), 2204.06411.

\bibitem{ATLAS:2017nyv}
ATLAS, M.~Aaboud {\em et~al.},
\newblock Phys. Lett. B {\bf 776}, 318 (2018), 1708.09624.

\bibitem{ATLAS:2018ojr}
ATLAS, M.~Aaboud {\em et~al.},
\newblock Eur. Phys. J. C {\bf 78}, 995 (2018), 1803.02762.

\bibitem{Ipek:2014gua}
S.~Ipek, D.~McKeen, and A.~E. Nelson,
\newblock Phys. Rev. {\bf D90}, 055021 (2014), 1404.3716.

\bibitem{No:2015xqa}
J.~M. No,
\newblock Phys. Rev. D {\bf 93}, 031701 (2016), 1509.01110.

\bibitem{Goncalves:2016iyg}
D.~Goncalves, P.~A.~N. Machado, and J.~M. No,
\newblock Phys. Rev. D {\bf 95}, 055027 (2017), 1611.04593.

\bibitem{Bauer:2017ota}
M.~Bauer, U.~Haisch, and F.~Kahlhoefer,
\newblock JHEP {\bf 05}, 138 (2017), 1701.07427.

\bibitem{Tunney:2017yfp}
P.~Tunney, J.~M. No, and M.~Fairbairn,
\newblock Phys. Rev. D {\bf 96}, 095020 (2017), 1705.09670.

\bibitem{Pani:2017qyd}
P.~Pani and G.~Polesello,
\newblock Phys. Dark Univ. {\bf 21}, 8 (2018), 1712.03874.

\bibitem{LHCDarkMatterWorkingGroup:2018ufk}
LHC Dark Matter Working Group, T.~Abe {\em et~al.},
\newblock Phys. Dark Univ. {\bf 27}, 100351 (2020), 1810.09420.

\bibitem{Haisch:2018kqx}
U.~Haisch, J.~F. Kamenik, A.~Malinauskas, and M.~Spira,
\newblock JHEP {\bf 03}, 178 (2018), 1802.02156.

\bibitem{Abe:2018emu}
T.~Abe, M.~Fujiwara, and J.~Hisano,
\newblock JHEP {\bf 02}, 028 (2019), 1810.01039.

\bibitem{Haisch:2018hbm}
U.~Haisch and G.~Polesello,
\newblock JHEP {\bf 02}, 128 (2019), 1812.08129.

\bibitem{Haisch:2018bby}
U.~Haisch and G.~Polesello,
\newblock JHEP {\bf 02}, 029 (2019), 1812.00694.

\bibitem{Abe:2019wjw}
T.~Abe, M.~Fujiwara, J.~Hisano, and Y.~Shoji,
\newblock JHEP {\bf 01}, 114 (2020), 1910.09771.

\bibitem{Butterworth:2020vnb}
J.~M. Butterworth, M.~Habedank, P.~Pani, and A.~Vaitkus,
\newblock SciPost Phys. Core {\bf 4}, 003 (2021), 2009.02220.

\bibitem{Arcadi:2020gge}
G.~Arcadi, G.~Busoni, T.~Hugle, and V.~T. Tenorth,
\newblock JHEP {\bf 06}, 098 (2020), 2001.10540.

\bibitem{Argyropoulos:2021sav}
S.~Argyropoulos, O.~Brandt, and U.~Haisch,
\newblock Symmetry {\bf 13}, 2406 (2021), 2109.13597.

\bibitem{Robens:2021lov}
T.~Robens,
\newblock Symmetry {\bf 13}, 2341 (2021), 2106.02962.

\bibitem{Arcadi:2022dmt}
G.~Arcadi and A.~Djouadi,
\newblock Phys. Rev. D {\bf 106}, 095008 (2022), 2204.08406.

\bibitem{Arcadi:2022lpp}
G.~Arcadi, N.~Benincasa, A.~Djouadi, and K.~Kannike,
\newblock Phys. Rev. D {\bf 108}, 055010 (2023), 2212.14788.

\bibitem{Haisch:2023rqs}
U.~Haisch and L.~Schnell,
\newblock JHEP {\bf 04}, 134 (2023), 2302.02735.

\bibitem{Argyropoulos:2024yxo}
S.~Argyropoulos, U.~Haisch, and I.~Kalaitzidou,
\newblock JHEP {\bf 07}, 263 (2024), 2404.05704.

\bibitem{chrisjayitame}
J.~Lahiri, K.~Rolbiecki, and T.~Robens,
\newblock {Constraining the Inert Doublet Model at the LHC},
\newblock To appear.

\bibitem{Drees:2013wra}
M.~Drees, H.~Dreiner, D.~Schmeier, J.~Tattersall, and J.~S. Kim,
\newblock Comput. Phys. Commun. {\bf 187}, 227 (2015), 1312.2591.

\bibitem{Dercks:2016npn}
D.~Dercks {\em et~al.},
\newblock Comput. Phys. Commun. {\bf 221}, 383 (2017), 1611.09856.

\bibitem{ATLAS:2021gcn}
ATLAS, G.~Aad {\em et~al.},
\newblock Phys. Lett. B {\bf 829}, 137066 (2022), 2111.08372.

\bibitem{EuropeanStrategyforParticlePhysicsPreparatoryGroup:2019qin}
R.~K. Ellis {\em et~al.},
\newblock (2019), 1910.11775.

\bibitem{p5rep}
{2023 Particle Physics Project Prioritization Panel},
\newblock {Exploring the Quantum Universe, Pathways to Innovation and Discovery
  in Particle Physics},
\newblock https://www.usparticlephysics.org/2023-p5-report/.

\bibitem{Alwall:2011uj}
J.~Alwall, M.~Herquet, F.~Maltoni, O.~Mattelaer, and T.~Stelzer,
\newblock JHEP {\bf 06}, 128 (2011), 1106.0522.

\bibitem{Robens:2024wbw}
T.~Robens,
\newblock EPJ Web Conf. {\bf 315}, 01025 (2024), 2409.19657.

\bibitem{deBlas:2024bmz}
J.~de~Blas {\em et~al.},
\newblock (2024), 2401.07564.

\bibitem{Gunion:1990kf}
J.~F. Gunion, H.~E. Haber, and J.~Wudka,
\newblock Phys. Rev. D {\bf 43}, 904 (1991).

\bibitem{Feuerstake:2024uxs}
F.~Feuerstake, E.~Fuchs, T.~Robens, and D.~Winterbottom,
\newblock JHEP {\bf 04}, 094 (2025), 2409.06651.

\bibitem{Bahl:2022igd}
H.~Bahl {\em et~al.},
\newblock Comput. Phys. Commun. {\bf 291}, 108803 (2023), 2210.09332.

\bibitem{Bechtle:2008jh}
P.~Bechtle, O.~Brein, S.~Heinemeyer, G.~Weiglein, and K.~E. Williams,
\newblock Comput. Phys. Commun. {\bf 181}, 138 (2010), 0811.4169.

\bibitem{Bechtle:2011sb}
P.~Bechtle, O.~Brein, S.~Heinemeyer, G.~Weiglein, and K.~E. Williams,
\newblock Comput. Phys. Commun. {\bf 182}, 2605 (2011), 1102.1898.

\bibitem{Bechtle:2013wla}
P.~Bechtle {\em et~al.},
\newblock Eur. Phys. J. C {\bf 74}, 2693 (2014), 1311.0055.

\bibitem{Bechtle:2020pkv}
P.~Bechtle {\em et~al.},
\newblock Eur. Phys. J. C {\bf 80}, 1211 (2020), 2006.06007.

\bibitem{Bechtle:2013xfa}
P.~Bechtle, S.~Heinemeyer, O.~St\r{a}l, T.~Stefaniak, and G.~Weiglein,
\newblock Eur. Phys. J. C {\bf 74}, 2711 (2014), 1305.1933.

\bibitem{Bechtle:2014ewa}
P.~Bechtle, S.~Heinemeyer, O.~St\r{a}l, T.~Stefaniak, and G.~Weiglein,
\newblock JHEP {\bf 11}, 039 (2014), 1403.1582.

\bibitem{Bechtle:2020uwn}
P.~Bechtle {\em et~al.},
\newblock Eur. Phys. J. C {\bf 81}, 145 (2021), 2012.09197.

\bibitem{filipcorfu}
A.~F. Zarnecki,
\newblock {Prospects for light exotic scalar measurements at the $e^+e^-$ Higgs
  factory},
\newblock These proceedings.

\bibitem{Black:2022cth}
K.~M. Black {\em et~al.},
\newblock JINST {\bf 19}, T02015 (2024), 2209.01318.

\bibitem{Accettura:2023ked}
C.~Accettura {\em et~al.},
\newblock Eur. Phys. J. C {\bf 83}, 864 (2023), 2303.08533,
\newblock [Erratum: Eur.Phys.J.C 84, 36 (2024)].

\bibitem{InternationalMuonCollider:2024jyv}
International Muon Collider, C.~Accettura {\em et~al.},
\newblock CERN Yellow Rep. Monogr. {\bf 2/2024}, 176 (2024), 2407.12450.

\bibitem{Dawson:1984gx}
S.~Dawson,
\newblock Nucl. Phys. B {\bf 249}, 42 (1985).

\bibitem{Kane:1984bb}
G.~L. Kane, W.~W. Repko, and W.~B. Rolnick,
\newblock Phys. Lett. B {\bf 148}, 367 (1984).

\bibitem{Chanowitz:1984ne}
M.~S. Chanowitz and M.~K. Gaillard,
\newblock Phys. Lett. B {\bf 142}, 85 (1984).

\bibitem{Gunion:1986gm}
J.~F. Gunion, J.~Kalinowski, and A.~Tofighi-Niaki,
\newblock Phys. Rev. Lett. {\bf 57}, 2351 (1986).

\bibitem{Chen:2016wkt}
J.~Chen, T.~Han, and B.~Tweedie,
\newblock JHEP {\bf 11}, 093 (2017), 1611.00788.

\bibitem{Kunszt:1987tk}
Z.~Kunszt and D.~E. Soper,
\newblock Nucl. Phys. B {\bf 296}, 253 (1988).

\bibitem{Costantini:2020stv}
A.~Costantini {\em et~al.},
\newblock JHEP {\bf 09}, 080 (2020), 2005.10289.

\bibitem{Han:2020uid}
T.~Han, Y.~Ma, and K.~Xie,
\newblock Phys. Rev. D {\bf 103}, L031301 (2021), 2007.14300.

\bibitem{Ruiz:2021tdt}
R.~Ruiz, A.~Costantini, F.~Maltoni, and O.~Mattelaer,
\newblock JHEP {\bf 06}, 114 (2022), 2111.02442.

\bibitem{Garosi:2023bvq}
F.~Garosi, D.~Marzocca, and S.~Trifinopoulos,
\newblock JHEP {\bf 09}, 107 (2023), 2303.16964.

\bibitem{Denner:2024yut}
A.~Denner and S.~Rode,
\newblock Eur. Phys. J. C {\bf 84}, 542 (2024), 2402.10503.

\bibitem{Cowan:2010js}
G.~Cowan, K.~Cranmer, E.~Gross, and O.~Vitells,
\newblock Eur. Phys. J. C {\bf 71}, 1554 (2011), 1007.1727,
\newblock [Erratum: Eur.Phys.J.C 73, 2501 (2013)].

\end{thebibliography}

\end{document}